\documentstyle[preprint,aps]{revtex}
\tightenlines

\begin{document}
\newcommand{\beq}{\begin{equation}}
\newcommand{\eeq}{\end{equation}}
\newcommand{\beqa}{\begin{eqnarray}}
\newcommand{\eeqa}{\end{eqnarray}}
\newcommand{\sr}{\sqrt}
\newcommand{\fr}{\frac}
\newcommand{\mn}{\mu \nu}
\newcommand{\G}{\Gamma}

\draft
\preprint{ INJE-TP-01-02}
\title{ Radially infalling brane and  moving domain wall \\in the brane cosmology }
\author{  Y.S. Myung\footnote{E-mail address:
ysmyung@physics.inje.ac.kr} }
\address{
Department of Physics, Graduate School, Inje University,
Kimhae 621-749, Korea}
\maketitle

\begin{abstract}

We discuss the brane cosmology in the 5D anti de Sitter
Schwarzschild (AdSS$_5$) spacetime. A brane with the
tension $\sigma$ is defined as the edge of an AdSS$_5$ space.
We  point out that the location of the horizon is an apparently, singular
point at where we may not define an embedding of the  AdSS$_5$
spacetime into the moving domain wall (MDW).
We resolve this problem  by introducing a radially
infalling brane (RIB) in AdSS$_5$ space, where an apparent
singularity turns out to be a coordinate one.
Hence the CFT/FRW-cosmology  is well-defined at the horizon.
As an example, an universal Cardy formula for the entropy of the  CFT
can  be given by the Friedmann equation at the horizon.
\end{abstract}

\newpage
\section{Introduction}
Recently there has been much interest in the phenomenon of
localization of gravity proposed by Randall and Sundrum
(RS)~\cite{RS1,RS2}.
RS assumed
a single positive tension 3-brane and a negative bulk cosmological
constant in the 5D spacetime~\cite{RS2}. They have obtained a 4D localized gravity
by fine-tuning the tension of the brane to the cosmological constant.
More recently, several authors have studied its cosmological
implications.
The brane cosmology   contains  some important
deviations from the Friedmann-Robertson-Walker (FRW) cosmology.
One  approach is first to assume the 5D dynamic metric (that is, BDL-metric\cite{BDL1,BDL2})
 which is
manifestly $Z_2$-symmetric.
Then one solves the  Einstein equation with a localized stress-energy tensor
to find the behavior of the scale factor.
 We call this   the BDL approach.

The other $Z_2$-symmetric  approach starts with
a static configuration which is two
AdSS$_5$ spaces joined by the domain wall.
 In this case  the
embedding into the moving domain wall\footnote{Here we use the term ``moving domain wall"
loosely to refer to any 3-brane moving in 5 dimensions.} is possible by choosing a
normal vector  $n_M$ and a tangent vector $u_M$\cite{CR,KRA,IDA}.
The domain wall separating two such bulk spaces is taken to be
located at $r=a(\tau)$, where $a(\tau)$ will  be determined by solving
the Israel junction condition\cite{ISR}. Then an observer on the wall will
interpret his motion through the static bulk background as
cosmological expansion or contraction\cite{CH}.

On the other hand, brane cosmology has been studied in the AdS/CFT correspondence.
For example, the holographic principle was investigated in a FRW
universe with a conformal field theory (CFT) within an AdSS$_5$-bulk theory\cite{VER}.
In this case the brane is considered as the  edge of an AdSS$_5$ space.
  The brane
starts with (big bang) inside the small black hole\footnote{
For the holographic entropy bound in cosmology, one may consider either a universe-size
black hole with $\ell=r_+$ or the large black hole with $\ell<r_+$\cite{VER}. But for  these
cases, one cannot choose an appropriate embedding for obtaining
 the moving domain wall (brane). Hence we
do not consider these black holes.} ($\ell>r_+$), crosses the horizon, and
expands until it reaches  maximum size. And then the brane contracts
, it falls the black hole again and finally disappears (big crunch). An
observer in  AdSS$_5$-space finds two interesting moments (two points in
the Penrose diagram\cite{MSM}) when the brane crosses the past (future) event
horizon. Authors in ref.\cite{SV} insisted that at these times the
Friedmann equation controlling the cosmological expansion (contraction)
coincides with an universal Cardy formula  for the entropy of the
CFT on the brane. If the above is true, it seems surprising that the Friedmann
equation contains information about thermodynamics of the CFT.
However, the Friedmann equation at the position of
the horizon is obscure
because the embedding of the AdSS$_5$ spacetime to the moving
domain wall  is apparently singular at these points.

In this paper, we resolve this  embedding problem  of an AdSS$_5$-black
hole spacetime into the moving domain wall  by introducing a radially
infalling brane (RIB), where the same problem occurs.
It turns out that in the case of RIB, an embedding onto the horizon (when the brane
crosses the black hole) can be defined because it belongs to a
coordinate singularity. Similarly we show that an embedding into
the MDW at the horizon is possible. We  can define the Friedmann equation at
these moments. Hence we  can introduce  a relation of the  CFT/FRW-cosmology on the brane.

For  cosmological embedding, let us start with an
AdSS$_5$-spacetime\cite{BIR} ,
\beq
ds^{2}_{5}= g_{MN}dx^Mdx^N= -h(r)dt^2 +\fr{1}{h(r)}dr^2 +r^2
\left[d\chi^2 +f_{k}(\chi)^2(d\theta^2+ \sin^2 \theta d\phi^2)
\right],
\label{BMT}
\eeq
where $k=0,\pm1$. $h(r)$ and $f_k(\chi)$ are given by
\beq
h(r)=k-\fr{m}{r^2}+ \fr{ r^2}{\ell^2},~~~
f_{0}(\chi) =\chi, ~f_{1}(\chi) =\sin \chi, ~f_{-1}(\chi) =\sinh
\chi.
\eeq
In the case of $m=0$, we have an exact AdS$_5$-space.  However,
$m \not=0$ generates the electric part of the Weyl tensor $E_{MP}=C_{MNPQ}n^Nn^Q$\cite{SMS}.
This  means that the bulk
spacetime has an  small black hole  ($\ell>r_+$)
 with  the horizon at $r=r_+$, $r_+^2= \ell^2(\sqrt{k^2 +4
m/\ell^2}-k)/2$\cite{KRA}. Hereafter we neither consider the
universe-size ($\ell=r_+)$ nor large ($r_+>\ell)$ black holes
because one cannot define their embedding into the moving domain wall\cite{VER}.

\section{Moving Domain Wall (MDW)}

Now we introduce the radial  location of a MDW in the form of
$ r=a(\tau),t=t(\tau)$ parametrized by the proper time $\tau$
: $(t,r,\chi,\theta,\phi)\to(t(\tau),a(\tau), \chi, \theta, \phi)$.
Then we expect that the induced metric of dynamical domain wall will be given
by the  FRW-type.
Hence $\tau$ and $a(\tau)$ will imply the cosmic time and scale factor of
the FRW-universe, respectively.
A tangent vector (proper velocity)  of this MDW

\beq
u= \dot t \fr{\partial}{\partial t}+ \dot a \fr{\partial}{\partial
a},\label{TAN}
\eeq
is introduced to define an embedding properly.
Here  overdots mean  differentiation with respect to
$\tau$. This is normalized to  satisfy
\beq
u^{M}u^{N} g_{MN}=-1.
\label{NTA}
\eeq
Given a tangent vector $u_M$, we need a normal 1-form
 directed toward to the bulk. Here we choose this as
\beq
n= \dot a dt-  \dot t da, ~~~n_{M}n_{N} g^{MN}=1.
\label{NNV}
\eeq
This convention  is consistent with the Randall-Sundrum case in the limit
of $m=0$\cite{RS2}. Using either Eq.(\ref{TAN}) with (\ref{NTA}) or
Eq.(\ref{NNV}), we can express the proper time rate of the AdSS$_5$
time
 $\dot t$ in terms of $\dot a$
as
\beq
\dot t=\fr{\sqrt{\dot a^2 +h(a)}}{h(a)}.
\label{TAV}
\eeq
From the above, it seems that $\dot t$ is not defined at $a=a_+$
because $h(a_+)=0$. This also happens in the study of static black hole.
Usually one introduces  a tortoise coordinate $r^*= \int h^{-1}dr$ to resolve it.
Then Eq.(\ref{BMT}) takes a form of $ds^2_5= -h(dt^2-dr^{*2})\cdots$ and
one finds the Kruscal extension. This means that $r=r_+$ is just a coordinate
singularity. We confirm this from $R_{MNPQ}R^{MNPQ}=40/\ell^4 + 72 m^2/r^8$,
which shows that $r=0$ $(r=r_+)$ are   true (coordinate) singularity.
Our dynamic situation is different from the static case.
Here a convenient coordinate is not a tortoise one $r^*$ but $r$
itself.
Eq.(\ref{TAV}) takes an
alternative form of
$h \dot t = \sqrt{\dot a^2 + h}$, which  implies that  $\sqrt{\dot a^2 + h}=0 \to
 \dot a=0$ at
$a=a_+$.
Let us study this point more carefully. An explicit form of our
 tangent vector is given by
\beq
u^M=\big(\fr{\sqrt{\dot a^2+ h(a)}}{h(a)}, \dot a,0,0,0 \big),~~~~
u_M=\big(-\sqrt{\dot a^2+ h(a)},\fr{ \dot a}{h(a)},0,0,0 \big).
\eeq
On the other hand, the normal vector takes the form
\beq
n^M=\big(-\fr{ \dot a}{h(a)},-\sqrt{\dot a^2+ h(a)},0,0,0 \big ),~~~~
n_M=\big(\dot a,-\fr{\sqrt{\dot a^2+ h(a)}}{h(a)},0,0,0 \big).
\eeq
As is emphasized again, two vectors which are essential for the
embedding are well defined everywhere, except $r=r_+$.
But these look like singular vectors at the horizon.
 Hence two moments when the brane crosses the past (future) event
horizons are singular points where one may not define the moving
domain wall. This persists in deriving the 4D intrinsic
metric and the extrinsic curvature.
The first two terms in Eq.(\ref{BMT}) together with Eq.(\ref{TAV}) leads to
\beq
-h(r)dt^2 + \fr{1}{h(r)} dr^2 ~~\to~~
-\big(h(a) \dot t^2 -\fr{\dot a^2}{h(a)} \big)d\tau^2= -d\tau^2.
\eeq
Here one may worry about this connection when $h(a_+)=0$.
The 4D induced line element is
\beqa
ds^{2}_{4}&&=-d \tau^2 +a(\tau)^2
\left[d\chi^2 +f_{k}(\chi)^2(d\theta^2+ \sin^2 \theta d\phi^2)
\right]
\nonumber \\
&&\equiv h_{\mu \nu}dx^{\mu} dx^{\nu},
\label{INM}
\eeqa
where we use the Greek indices only  for the brane.
Actually the embedding of an AdSS$_5$ space to the FRW-universe is a
$2(t,r) \to 1(\tau)$-mapping. The  projection tensor is given by $h_{MN}=g_{MN}-n_Mn_N$
and its determinant is zero. Hence its inverse  $h^{MN}$ cannot be
defined. This means that the above embedding belongs to a peculiar
mapping to obtain the induced metric $h_{\mu\nu}$ from the AdSS$_5$ black
hole spacetime $g_{MN}$ with $n_M$.
In addition, the extrinsic curvature is defined  by

\beqa
&&K_{\tau\tau}=K_{MN} u^M u^N =(h(a) \dot t)^{-1}(\ddot a +h'(a)
 /2)=\fr{\ddot a +h'(a)/2}
{\sqrt{\dot a^2 +h(a)}}, \\
&&K_{\chi\chi} = K_{\theta\theta}=K_{\phi\phi}
=- h(a) \dot t a=-\sqrt{\dot a^2 +h(a)}~a,
\eeqa
where  prime stands for  derivative with respect to $a$.
We observe that $K_{\tau\tau}$ looks like ill-defined as $\fr{\ddot a +h'(a)/2}{0}$, and
 $K_{\chi\chi} = K_{\theta\theta}=K_{\phi\phi}=0$ at $a=a_+$.
As we will see later,  this belongs to  apparent phenomena.
A localized matter on the brane implies that the extrinsic curvature jumps  across the brane.
This jump is  described  by the Israel junction condition

\beq
K_{\mu \nu}=-\kappa^2 \left(
T_{\mu\nu}-\fr{1}{3}T^{\lambda}_{\lambda}h_{\mu\nu} \right)
\label{4DI}
\eeq
with $\kappa^2=8 \pi G_5^N$.  For cosmological purpose,
we may  introduce  a localized stress-energy tensor on the
brane as the 4D perfect fluid

\beq
T_{\mu \nu}=(\varrho +p)u_{\mu}u_{\nu}+p\:h_{\mu\nu}.
\label{MAT}
\eeq
Here $\varrho=\rho+ \sigma$ $(p=P-\sigma)$, where $\rho $ $(P)$ is the energy density (pressure)
of the localized matter and $\sigma$ is the brane tension.
In the case of $\rho=P=0$, the r.h.s. of
Eq.(\ref{4DI}) leads to  a form of the RS case as $-\fr{\sigma \kappa^2}{3}
h_{\mu\nu}$.
From Eqs.(\ref{4DI}), one finds
 the space component of the junction condition

\beq
\sqrt{h(a) + \dot a^2}=\fr{\kappa^2}{3}\sigma a.
\label{SEE}
\eeq
For a single AdSS$_5$, we have the fine-tuned brane tension $\sigma=3/(\kappa^2\ell)$.
The above equation   leads to
\beq
H^2=- \fr{k}{a^2} +\fr{m}{a^4},
\label{HHH}
\eeq
where $H=\dot a/a$ is the expansion rate. The term of $m/a^4$
originates from the electric (Coulomb) part of the 5D Weyl tensor, $E_{00} \sim
m/a^4$\cite{SMS,MSM}. This term behaves like radiation\cite{BDL2}.
Especially for $k=1$, we have  $m=\fr{16 \pi G_5^N M}{3 V(S^3)},M=\fr{a}{\ell}E,
 V=a^3 V(S^3), G_5^N=\fr{\ell}{2} G_4^N$. Then one finds a CFT-radiation dominated
universe
\beq
H^2=- \fr{1}{a^2} +\fr{8\pi
G_4^N}{3}\rho_{CFT},~~~\rho_{CFT}=\fr{E}{V}.
\label{CRA}
\eeq
It seems  that the equation (\ref{SEE}) is well-defined even at
$a=a_+$. Thus this leads to $H= \pm 1/\ell$ at $a=a_+$, which
is just the case mentioned  in ref.\cite{SV}. At this stage, this point is not clear.
 From
Eq.(\ref{TAV}), one finds $\dot a^2=0$ at $a=a_+$. This means that
$H^2a^2=0 \to H=0$ at $a=a_+$ because of $a_+ \not=0$. Naively we find
a contradiction.  Also, considering the extrinsic
curvature expressed in terms of $a,h(a)$, the junction condition
may not be defined at $a=a_+$. To resolve this problem, we
introduce a radially infalling brane, where the same
situation occurs as in the MDW picture.

\section{Radially Infalling Brane (RIB)}
A 3-brane action is usually given by the Nambu-Goto action. Here
for cosmological purpose, we consider its point particle limit of
brane $\to$ body. The corresponding action with unit mass
on the AdSS$_5$ background space\cite{CHAN} is
given by
\beq
{\cal L}=- \fr{1}{2} g_{MN}\fr{dx^M}{d\tau}\fr{dx^N}{d\tau}
=\fr{1}{2}\big( h \dot t^2 -\fr{\dot a^2}{h}+ \cdots \big)
\label{RIF}
\eeq
where $\cdots$ means the angular part. This part is not relevant to our
purpose because we consider only  radial time-like geodesics.

For time-like geodesics, $\tau$ is  proper time of  the RIB
describing the geodesic. The corresponding canonical momenta are
\beq
p_t= \fr{\partial {\cal L}}{\partial \dot t}= h(a)\dot t
,~~~ p_a=-\fr{\partial {\cal L}}{\partial \dot a}= \fr{\dot
a}{h(a)}, \cdots.
\label{CAM}
\eeq
$p_t$ $(p_a)$ correspond to $-u_t=-n^a$ $(u_a=-n^t)$ in the MDW approach.
Hence the apparent singularity at $a=a_+$ appears in the RIF picture.
We always choose $2{\cal L}=1$ by rescaling the affine
parameter $\tau$ for time-like geodesics.
This is just the same as was found in the normalization condition
Eq.(\ref{TAV}) for the tangent vector $u^M$ and normal vector
 $n_M$ in the MDW approach. Here we get an integral of the motion
from  the fact that $t$ is a cyclic coordinate :
$\fr{dp_t}{d \tau}=\fr{\partial {\cal L}}{\partial t}=0$,
\beq
p_t= h(a)\dot t= \sqrt{ \dot a^2 +h(a)}= E
\label{INTG}
\eeq
where $E$ is a constant of the motion.
The above equation corresponds to  Eq.(\ref{SEE})  in the MDW approach.
 On the other hand, $2{\cal L}=1$
means
\beq
\fr{1}{h(a)}( E^2- \dot a^2)= 1.
\label{LAN}
\eeq
This is nothing new and corresponds to Eq.(\ref{TAV}) in the MDW approach.
Different choices of the constant $E$ corresponds to different
initial conditions. For simplicity, let us make a choice of $E=1$, which
corresponds to dropping in a RIB from infinity with  zero initial
velocity. In the limit of $\ell \to \infty$, the situation
becomes rather clear. Hence first we consider the geodesic of RIB in the
5D Schwarzschild  black hole spacetime.

\subsection{RIB in 5D Schwarzschild space}

In the case of  $\ell \to \infty$, we find from
Eq.(\ref{LAN})
\beq
\fr{1}{\dot a^2}= \fr{a^2}{m} \to ~~~d\tau = -\fr{a}{\sqrt{m}}da,
\label{ERIB}
\eeq
where we take the negative square root because we consider an
infalling  body into the black hole from the large $a_0$ (
$a_0>> \sqrt{m}$).
This leads to
\beq
\tau-\tau_0 = \fr{1}{2\sqrt{m}}( a^2_0-a^2)
\eeq
where the RIB is located at $a_0$ at proper time $\tau_0$. Any singular
behavior does not appear at the Schwarzschild radius $a=a_+=\sqrt{m}$
and the RIB falls continuously to $a=0$ in a finite proper time.
But if we describe the motion of RIB in terms of the
Schwarzschild coordinates $(t,a)$, then
\beq
\fr{dt}{da}=\fr{ \dot t}{\dot a}= -\fr{a}{\sqrt{m} h(a)}
\eeq
which is integrated as
\beq
t-t_0= \fr{1}{2\sqrt{m}} \big[ a_0^2 -a^2 + m
\log\fr{(a_0-\sqrt{m})(a_0+\sqrt{m})}{(a-\sqrt{m})(a+\sqrt{m})}\big].
\eeq
In the limit of $a \to a_+=\sqrt{m}$, one has
\beq
t-t_0 \simeq -\fr{\sqrt{m}}{2} \log (a-\sqrt{m})\to \infty
\eeq
which means that $a_+=\sqrt{m}$ is approached but never passed.
The coordinate $t$ is useful and physically meaningful
asymptotically at large $a$ since it corresponds to proper time
measured by an observer at rest far away from the origin (that
is, $dt=d\tau$ when $a \to \infty$).  From the point of view of
such an observer, it takes an infinite time for the RIB to reach
$a=a_+$.
On the other hand, from the point of view of the RIB itself,
it reaches $a=a_+$ and $a=0$ within  finite proper time. Clearly,
the Schwarzschild time coordinate $t$ is inappropriate for
describing  a radially infalling motion.

\subsection{RIB in AdSS$_5$ space}

Now we are in a position to study the motion of RIB in the AdSS$_5$
black hole spacetime. In this case, we have
\beq
\fr{1}{\dot a^2}= \fr{a^2\ell^2}{m\ell^2-a^4} \to ~~~d\tau = -\fr{a\ell}
{\sqrt{m\ell^2-a^4}}da
\label{ARIB}
\eeq
which leads to
\beq
\tau-\tau_0 =\fr{\ell}{2}
\big(\tan^{-1}\big[\fr{a_0^2}{\sqrt{m\ell^2 -a_0^4}}\big]-
\tan^{-1}\big[\fr{a^2}{\sqrt{m\ell^2 -a^4}}\big]\big)\leq \fr{\pi
\ell}{4}.
\eeq
Hence it takes  finite proper time for a RIB to reach $a=a_+$ and
$a=0$. In order to a simple relation between $\tau$ and $t$, let us
consider the asymptotic form of  AdSS$_5$ space with
$k=1$ :$\lim_{a \to \infty}[\fr{\ell^2}{a^2}ds_5^2]=-dt^2 +\ell^2 d\Omega^2_3.$
We  find that the proper time $\tau$  is equal to the AdSS$_5$ time $t$
only when the radius of $S^3$ is set to be $\ell$.
 For a finite $a$,
 the relation becomes quite complicated. Here we have
\beq
\fr{dt}{da}=\fr{ \dot t}{\dot a}= -\fr{a\ell}{\sqrt{m\ell^2-a^4}}\fr{1}{h(a)}.
\label{DTDA}
\eeq
We note that integration of  Eq.(\ref{DTDA}) leads to a complicated from for
$t-t_0$. This is  related to  the presence of a term  $a^2/\ell^2$ in $h(a)$.
Hence the proper time $\tau$ is an affine variable which describes
 time-like geodesic (the motion of RIB) correctly in  AdSS$_5$ space.
On the other hand, it turns out that the  AdSS$_5$ time coordinate $t$
is not appropriate for describing the RIB which falls into the
Schwarzschild black hole in anti de Sitter space.

\section{Discussion}
First let us compare the MDW with the RIB.
The equation of  MDW  with $k=1$ is given by
\beq
\fr{1}{2} \dot a^2 +V(a)_{MDW} =-\fr{1}{2}
\eeq
with  its potential $V(a)_{MDW}=- \fr{m}{2 a^2}$.
This corresponds to the motion of a point particle with unit mass
and a negative energy rolling in a  potential $V_{MDW}$. Actually
the scale factor  increases from $a=0$ to maximum size, $V(a_{max})_{MDW}=-1/2$,
and recollapses into $a=0$.
On the other hand, the motion of  RIB is given by
\beq
\fr{1}{2} \dot a^2 +V(a)_{RIB} =0
\eeq
with  a potential $V(a)_{RIB}=- \fr{m}{2 a^2}+ \fr{a^2}{2\ell^2}$.
This corresponds to the equation for a particle of unit mass and
zero energy rolling in a potential $V_{RIB}$. Hence it corresponds
to a radially infalling body starting at $a_0>> a_+$.
Hence it is  suggested that the motion of  RIF
covers half that of  MDW. For $k=0$ case, the MDW takes the same
equation as in the RIB  with a slightly different potential.
But this does not make a significant change. In this case, the
MDW starts at $a=0$ and ends up at $a=\infty$\cite{MSM}. As a
reverse process, the RIB starts at $a=\infty$ and ends up at
$a=0$.
Furthermore, at the horizons, the expansion rate of MDW is
 $H_{MDW}=\pm1/\ell$ whereas that of  RIB is
$H_{RIB}=\pm1/a_+$.

However an important difference is that
$H_{RIB}^2=-1/\ell^2 + m/a^4$ is not a kind of Friedmann-like equations.
The reason is that as is shown in $H_{RIB}^2=\cdots$, the first
term of $1/\ell^2$ means that the size of the background AdSS$_5$ space where the RIB
moves is always
fixed as the spatial curvature of the AdSS$_5$ ($\ell$).
On the other hand, the Friedmann-like equation (\ref{CRA}) for the MDW means
that at each instant the size of universe is given by the spatial curvature
(scale factor :$a$).

Anyway, it is clear from the analysis of the RIB
that   a genuine coordinate for the MDW is not the AdSS$_5$ time coordinate $t$
but the proper time $\tau$. Finally it turns out that   the apparent
singular behaviors of the normal, tangent vectors, and extrinsic
curvature at the horizon belong to coordinate artifacts.
Hence Eq.(\ref{CRA}) implies a radiation-dominated
universe of $\rho \sim a^{-4}$. This radiation can be identified
with the finite temperature CFT that is dual to the AdSS$_5$
geometry. This prescription is valid when the MDW crosses the
horizons. Thanks to this , an universal Cardy formula for the entropy of the  CFT
can  be given by the Friedmann equation at the horizon\cite{SV}.

In conclusion, the moving domain wall approach provides us  a nice tool to study
 the brane cosmology including the location of the horizon.

\section*{Acknowledgments}

This work was supported in part by the Brain Korea 21 Program, Ministry of
Education, Project No. D-0025 and KOSEF, Project No. 2000-1-11200-001-3.

\end{document}